\documentclass[journal]{IEEEtran}

\usepackage{graphicx}
\usepackage{latexsym}
\usepackage{amsfonts}
\usepackage{amssymb}
\usepackage{amsbsy}
\usepackage{amsmath}
\usepackage{multicol}
\usepackage{float}
\usepackage{subfigure}
\usepackage{algorithm}
\usepackage[dvips]{epsfig}

\begin{document}

\title{\LARGE Worst-case User Analysis in Poisson Voronoi Cells}

\author{Sang Yeob Jung, Hyun-kwan Lee, and Seong-Lyun Kim
\thanks{This research was supported by the International Research \&
Development Program of the National Research Foundation of Korea
(NRF) funded by the Ministry of Education, Science and
Technology (MEST) of Korea (Grant number: 2012K1A3A1A26034281, FY
2012).}
\thanks{The authors are with the Radio Resource Management $\&$ Optimization
Laboratory, Department of Electrical and Electronic Engineering,
Yonsei University, 50 Yonsei-Ro, Seodaemun-Gu, Seoul 120-749, Korea.
Email: \{syjung, hklee, slkim\}@ramo.yonsei.ac.kr.}}

\maketitle

\begin{abstract}
In this letter, we focus on the performance of a worst-case
mobile user (MU) in the downlink cellular network. We derive the coverage
probability and the spectral efficiency of the worst-case MU using stochastic geometry.
Through analytical and numerical results, we draw out interesting
insights that the coverage probability and the spectral efficiency of the worst-case MU decrease down to 23\% and 19\% of those of a typical MU, respectively.
By applying a coordinated scheduling (CS) scheme, we also investigate how much the performance of the worst-case MU is improved.
\end{abstract}


\section{Introduction}
One of the challenges facing the next-generation
wireless networks is to cope with the expected demand for data. Not
only to support high quality of service (QoS) but also to improve
spectral efficiency, orthogonal frequency division multiplexing (OFDM)-based cellular systems
have been widely deployed. An essential requirement for OFDM-based
cellular systems, however, is to specify and to enhance the
performance of cell-edge mobile users (MUs) or worst-case MUs
\cite{gpp_10}. Unlike the code division multiple access (CDMA) cellular system
that is robust against interference, the OFDM-based cellular network
suffers from high inter-cell interference (ICI) at the cell
boundary, especially due to the smaller cell size and denser reuse of spectrum in future mobile communications.

There has been a substantial amount of work on evaluating the performance of
cell-edge MUs or worst-case MUs on a regular hexagonal model
\cite{Tse_05}-\cite{slkim}. However, this model is highly idealized
and less accurate as the base station (BS) can be randomly
located to support a large number of MUs. To reflect
the actual BS deployment, cellular networks are recently modeled using stochastic geometry \cite{Andrews_11}--\cite{Dhillon_13}.
The coverage probability and the average ergodic rate of
\textit{a typical MU}\footnote{The term a typical MU means a randomly chosen MU over the entire network
so that the main focus is on the \textit{average} performance in the network.}
are derived \cite{Andrews_11} and load distribution is studied by considering the user density \cite{SMYU_13}.
By pushing the typical MU into the cell interior with the concept of conditional thinning,
the non-uniform MU distribution model is analyzed in \cite{Dhillon_13}
but all the results in \cite{Andrews_11}-\cite{Dhillon_13} are not about the performance of cell-edge MUs or worst-case MUs.
Since the performance of worst-case MUs was estimated almost exclusively via simulation or field trial over a few representative scenarios \cite{gpp_10}, \cite{R_Irmer_11}
that are time-consuming and difficult to apply in the general cellular networks,
it is crucial to provide a general theoretical analysis of the performance of the worst-case MUs.

The main purpose of this letter is to serve as a baseline for future
research on the performance of the worst-case MU. Modeling the
location of the BSs by a Poisson point process (PPP),
we address the performance of the worst-case MU analytically.
Under i.i.d. Rayleigh fading on all links, we derive the coverage
probability and the spectral efficiency.


To get insights on the performance of the worst-case MU, we compare our analytical
results with those for the typical MU derived by Andrews \textit{et al} \cite{Andrews_11}. Our conclusion is that the coverage probability of the worst-case MU decreases down to 23\% of that of the typical MU
at the target signal-to-interference-ratio (SIR) $\gamma=-2$ dB.
The spectral efficiency of the worst-case MU becomes 19\% of that of the typical MU.
Additionally, we investigate how much the performance of the worst-case MU is improved by applying a coordinated scheduling (CS) scheme \cite{gpp_10}.

\section{System Model}

\begin{figure}[t]
\centering
\includegraphics[angle=0,width=2.6in]{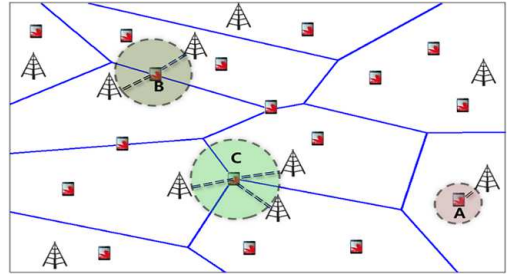}
\caption{Three types of mobile users. Point A
is an inner-cell MU with one neighbor BS, point B is an edge-cell MU with
two neighbor BSs, and point C is a worst-case MU with three neighbor
BSs.} \label{Fig1}
\end{figure}

Consider the downlink cellular network consisting of
BSs and MUs. We assume the locations of the BSs by
points of a homogeneous PPP $\Phi$ on
the plane with intensity $\lambda$. If an MU is associated with the nearest BS,
then the resulting coverage areas are divided into Voronoi cells and
such a partition is called a \textit{Poisson Voronoi tessellation}
\cite{Stoyan_96}.

\subsection{Three types of mobile users}
Depending on how many nearest BSs to a given MU, we can define three
types of the MUs in the network: An \textit{inner-cell} MU is in the Voronoi cell with
exactly one closest BS. An
\textit{edge-cell} MU with two closest BSs is on the boundary between
two Voronoi cells, and a \textit{worst-case} MU with three closest BSs
is in a vertex, where three Voronoi cells meet, as shown in Fig.
\ref{Fig1}. There are no MUs with four or more closest BSs with respect to a homogeneous PPP \cite{Stoyan_96}.

\subsection{Downlink SINR for worst-case MUs}
For the sake of simplicity and tractability, we restrict ourselves
to the following assumptions:
\begin{itemize}
\item Each BS transmits a constant power of $1/\mu$.
\item The fading on all links is i.i.d. Rayleigh distributed with mean $1$.
\item An MU can be served by at most one BS.
\item No intra-cell interference is present due to orthgonal frequency division multiple access (OFDMA).
\item The noise power at any MU receiver is $\sigma^{2}$.
\end{itemize}

To investigate the signal-to-interference-plus-noise-ratio (SINR) for a worst-case MU,
we assume that it is served by one of the nearest BSs.
The authors in \cite{Stoyan_96} showed
that the worst-case MUs still follow a homogeneous Poisson point process
with intensity $2\lambda$.

Without loss of generality, we denote the serving BS for the worst-case MU by
$b_{0}$ (i.e., the other two BSs
having the same distance from the worst-case MU are denoted
by $b_{1}$ and $b_{2}$, respectively). Assuming that the worst-case MU is located at the origin, the resulting SINR at
its associated BS at a random distance $r$ is given by
\begin{eqnarray}
\Gamma  = {{g_0 r^{ - \alpha } } \over {\sigma ^2  + I_r }},
\end{eqnarray}
where
\begin{eqnarray*}
I_r  = g_1 r^{ - \alpha}  + g_2 r^{ - \alpha }  + \sum\limits_{i \in \Phi \backslash \{ b_0 ,b_1 ,b_2 \} } {g_i R_i ^{ - \alpha } }.
\end{eqnarray*}
$I_r$ is the total received interference at the location of the worst-case MU from all BSs other than its associated BS.
Note that there are two nearest interfering BSs located at a distance $r$ and the other interfering BSs from the distance $R_{i}$
to the worst-case MU. The value $g_i$ follows an exponential distribution.

\section{Coverage Probability}
Let us derive the \textit{coverage} probability of a
worst-case MU in the downlink cellular network. The coverage
probability is defined by
\begin{eqnarray}
p_c (\gamma ,\lambda ,\alpha ) \buildrel \Delta \over =
\mathbb{P}[\Gamma  > \gamma ].
\end{eqnarray}

\subsection{Distributional properties at worst-case MUs}
Recall that the worst-case MU has three equidistant and nearest BSs.
Based on the theory of Palm distributions, it is known
in \cite{Muche_05} that the probability density function (PDF) of $r$ of (1) is given by
\begin{eqnarray}
f_r (r) = 2(\lambda \pi )^2 r^3 e^{ - \lambda \pi r^2 }.
\end{eqnarray}
Since a typical MU is an inner-cell MU,
its PDF $2\lambda\pi  re^{ - \lambda \pi r^2 }$ given in \cite{Andrews_11} is different from (3).
Using integration by parts, the null probability of the worst-case MU is
$\mathbb{P}[r > R] = \left( {1 + \lambda \pi R^2 } \right)e^{ -\lambda \pi R^2 }$ and its cumulative density function (CDF) is
$\mathbb{P}[r\leq R]=1-(1+\lambda\pi R^{2})e^{-\lambda\pi R^{2}}$.

\subsection{Main Result}
We now explain our main result for the coverage probability of the worst-case MU under i.i.d. Rayleigh fading.

\vskip 10pt \noindent {\bf Proposition 1.} {\it Assuming i.i.d.
Rayleigh fading on all links, the coverage probability of
a worst-case MU located at a typical vertex is:
\setlength\arraycolsep{3.5pt}
\begin{equation}\label{Proposition1}
p_c (\gamma, \lambda, \alpha)=2\left( {{{\lambda \pi } \over {1 +
\gamma }}} \right)^2 \int_0^\infty  {e^{ - \lambda \pi r^2 \left( {1
+ \rho \left( {\gamma ,\alpha } \right)} \right) - \mu \gamma
r^\alpha \sigma ^2 } r^3 } dr,
\end{equation}}
\textit{where}
\begin{eqnarray*}
\rho \left( {\gamma ,\alpha } \right) = \gamma ^{2/\alpha }
\int_{\gamma ^{ - 2/\alpha } }^\infty  {{1 \over {1 + u^{\alpha /2}
}}du.}
\end{eqnarray*}
\noindent {\bf Proof.} Appendix A. \hfill $\blacksquare$ \vskip 10pt

In the \textit{interference-limited} system, i.e., the noise power $\sigma^{2}$ is negligible compared to the total received interference,
the probability (4) can be simplified to
\begin{eqnarray}
p_c (\gamma ,\lambda ,\alpha) = \left( {{1 \over {\left( {1 + \gamma }
\right) {(1+\rho(\gamma,\alpha))}}}} \right)^2.
\end{eqnarray}
If the path loss exponent $\alpha=4$, the probability (5) can be further reduced to the following simple-closed form
\begin{eqnarray}
p_c (\gamma ,\lambda ,4) = \left( {{1 \over {\left( {1 + \gamma } \right)\kappa (\gamma )}}} \right)^2,
\end{eqnarray} \noindent
where $\kappa (\gamma ) = 1 + \rho (\gamma ,4) = 1 + \sqrt \gamma  \left( {\pi /2 - \tan ^{ - 1} \left( {1/\sqrt \gamma  } \right)} \right)$.

An interesting observation is that the coverage probability of the worst-case MU is independent of the BS density $\lambda$.
Even if more BSs are deployed, this does not affect the coverage probability of the worst-case MU since the received and interference powers
cancel each other out \cite{Andrews_11}.
Thus, some advanced techniques are needed to increase it.
It is noticeable that the coverage probability of the typical MU is $1/\kappa(\gamma)$ \cite{Andrews_11}, which is
higher than (6).

\subsection{Coordinated scheduling scheme}
A coordinated scheduling (CS) scheme is considered as a promising technique to mitigate ICI \cite{gpp_10}.
Since coordinated BSs can share channel information from the MU,
only one BS that has the best channel condition from the MU transmits to the MU to guarantee its performance in a given subframe.
To see the impact of the CS scheme, we focus on the interference-limited system with $\alpha=4$.
Then, the coverage probability is given by
\vskip 10pt \noindent {\bf
Corollary  1.}{\it\; The coverage probability of the worst-case MU in the downlink coordinated scheduling scheme is:
\begin{equation}
p_c ^{CS} \left( {\gamma ,\lambda ,4} \right) = {3 \over {\kappa \left( \gamma  \right)^2 }} - {3 \over {\kappa \left( {2\gamma } \right)^2 }} + {1 \over {\kappa \left( {3\gamma } \right)^2 }}.
\end{equation}}
\noindent {\bf Proof.} Appendix B.\hfill $\blacksquare$ \vskip 10pt
Comparing with (6), we can see that the better coverage probability is
achieved as shown in \textrm{F}ig. 2. A question is how much the probability is improved.
In the next subsection,
we compare (7) with $1/\kappa(\gamma)$, that of the typical MU.

\subsection{Coverage comparison}
\begin{figure}[t]
\centering
\includegraphics[angle=0,width=3.8in]{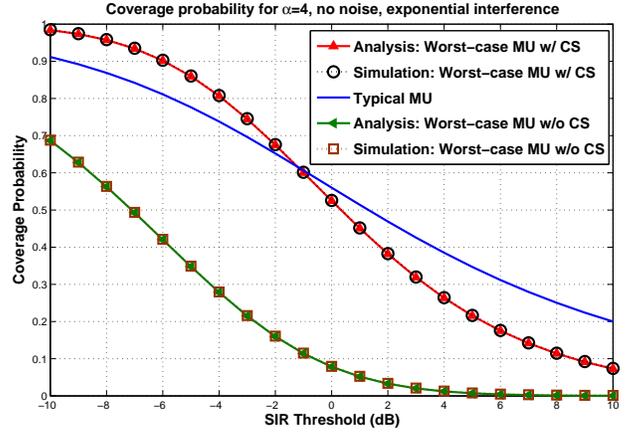}
\caption{Coverage probability as a function of the target SIR, $\gamma$. The BS density $\lambda$ does not affect the probability.} \label{Fig2}
\end{figure}

\textrm{F}ig. \ref{Fig2} shows the coverage probability versus the SIR threshold $\gamma$ under i.i.d. Rayleigh fading
on all links with $\alpha=4$. As noted, the coverage probability of the worst-case MU is lower than that of the typical MU \footnote{ We numerically obtained that
the coverage probability of an edge-cell MU is approximately 10\% higher than that of a worst-case MU on average.}.
Since the received signal decays exponentially with the propagation distance,
the worst-case MU experiences large path loss, which decreases the coverage probability of the worst-case MU down to 23\% of that of the typical MU at $\gamma=-2$\;dB.
Given that, by applying the CS scheme,
the better coverage probability can be achieved.
In terms of the coverage probability, there is a cross-over point between the worst-case MU with the CS scheme and the typical MU
at $\gamma=-1$\;dB.
Below the cross-over point, the enhanced signal quality
completely pay off the large path loss, which leads to a better coverage probability compared with that
of the typical MU. Beyond the cross-over point, however, the large path loss prevails over the increased signal strength,
which causes the coverage probability to decrease dramatically beyond the cross-over point.
Overall, the CS scheme improves the coverage probability of the worst-case MU, virtually shifting it to a typical MU.

\section{Spectral Efficiency}
In this section, we derive the spectral efficiency of a
worst-case MU and investigate how much it is improved by the CS scheme.\vskip
10pt \noindent {\bf Proposition 2.} {\it In the interference-limited system with $\alpha=4$, the spectral efficiency of a
worst-case MU located a typical vertex is:
\begin{eqnarray}
\tau_c(\lambda, 4)=\int_{t > 0}^{} {\left( {{{e^{ - t} } \over { \varepsilon_c (t)}}} \right)} ^{\rm{2}} {\rm{ }}dt,
\end{eqnarray}
where
\begin{eqnarray}
\varepsilon_c (t) =1+ \sqrt {e^t  - 1} \left( {\pi /2 - tan^{-1} \left( {1/\sqrt {e^t  - 1} } \right)} \right). \nonumber
\end{eqnarray}
}
\noindent {\bf Proof.} Appendix C. \hfill $\blacksquare$ \vskip 10pt \noindent
The spectral efficiency (8) can be evaluated numerically, which gives
\begin{eqnarray}
\tau_c (\lambda ,4) \approx {\rm{0.27nats/sec/Hz}} = {\rm{0.39\;bps/Hz}}.
\end{eqnarray}

The spectral efficiency (9) is 19\% of that of the typical MU, 2.15 bps/Hz derived in \cite{Andrews_11}.
By applying the CS scheme, we have the following.
\vskip
10pt \noindent {\bf Corollary 2.} {\it The spectral efficiency of the worst-case MU in the downlink coordinated scheduling scheme is:
\setlength\arraycolsep{3pt}
\medmuskip=-1mu
\begin{equation}
\tau _c^{CS} (\lambda ,4) = \int_{t > 0}^{} {{3 \over {(\varepsilon_c ^{CS} (1,t))^2 }}}  + {1 \over {(\varepsilon_c ^{CS} (3,t))^2 }} - {3 \over {(\varepsilon_c ^{CS} (2,t))^2 }}dt,\\
\end{equation}
where
\setlength\arraycolsep{-1pt}
\begin{eqnarray}
\varepsilon _c ^{CS} (a,t) =1+ \sqrt {a(e^t  - 1)} \left( {\pi /2 - \tan ^{ - 1} \left( {1/\sqrt {a(e^t  - 1)} } \right)} \right)
.\nonumber
\end{eqnarray}
}
\noindent {\bf Proof.} Appendix D \hfill $\blacksquare$ \vskip 10pt \noindent
By computing numerically, the spectral efficiency in the CS scheme is:
\begin{eqnarray}
\tau_c ^{CS} \left( {\lambda ,4} \right) \approx {\rm{1.49\;bps/Hz}},
\end{eqnarray} \vskip 10pt \noindent
where we see that the CS scheme has a profound effect on the spectral efficiency of the worst-case MU, which
becomes 69\% of that of the typical MU. Significant performance gain can be achieved for the worst-case MU.
On the other hand, there are some performance degradations in practice due to the latency for coordination \cite{R_Irmer_11}. Also,
the area spectral efficiency might be decreased by muting two dominant interferers.

\section{Conclusion}

In this letter, we derive the coverage probability and the spectral efficiency of the worst-case MU in the donwnlink cellular network.
To the best of our knowledge, this is the first approach to analytically derive the performance of the worst-case MU, using the stochastic geometry.
Even if simplifying assumptions are made, we believe our analysis
plays vital roles in analyzing randomly deployed BSs and serving as a reference for future research on the performance of the worst-case MU.

\section*{Appendix}

\subsection{Proof of Proposition 1}
Assuming that the worst-case MU is associated with one of
the nearest BSs at a distance $r$, the coverage probability is
\setlength\arraycolsep{1pt}
\begin{eqnarray}
p_c (\gamma ,\lambda ,\alpha) &\mathop=& \mathbb{E}_r \left[
{\mathbb{P}\left[ {\Gamma  >
\gamma |r} \right]} \right]\nonumber\\
&\mathop=&\int_{r > 0}^{} {\mathbb{P}\left[ {{{g_0 r^{ - \alpha } }
\over {\sigma ^2  + I_r }} > \gamma |r} \right]} {\rm{ }}f_r
(r)dr\nonumber\\
&\mathop=& \int_{r>0}  {\mathbb{E}_{I_r } \left[
{\mathbb{P}\left[ {g_0 > \gamma r^\alpha \left( {\sigma ^2  + I_r }
\right)|r,I_r } \right]}
\right]} {\rm{ }}f_r (r)dr \nonumber\\
&\mathop  = \limits^{(a)}& \int_{r>0}  {\mathbb{E}_{I_r } \left[
{e^{ - \mu \gamma r^\alpha  \left( {\sigma ^2  + I_r } \right)} |r}
\right]}
{\rm{ }}f_r (r)dr \nonumber\\
&\mathop=& 2\lambda ^2 \pi ^2 \int_{r>0}  {e^{ - \mu \gamma
r^\alpha  \sigma ^2 } \mathcal{L}_{I_r } \left( {\mu \gamma r^\alpha
} \right)} {\rm{ }}r^3 e^{ - \lambda \pi r^2 } dr, \nonumber\\
\end{eqnarray}
where $(a)$ follows from the fact that the random variable $g_0$ is
exponentially distributed with mean $1/\mu$. Note that $\mathcal{L}_{I_r}(.)$ is the
Laplace Transform of the cumulative interference from the all interfering BSs to
the worst-case MU and the distribution
$f_{r}(r)$ is given in \textrm{S}ubsection III-A. Using the stationarity of PPP and the independence of the fading random variables $\mathcal{L}_{I_r } (s)$ gives
\setlength\arraycolsep{-0.5pt}
\begin{eqnarray}
\mathcal{L}_{I_r } (s) &\mathop=& \mathbb{E}_{I_r } \left[ {e^{ - sI_r } } \right]
\mathop= \mathbb{E}_{\Phi ,g_i } \left[ {\exp\left( { - s\sum\limits_{i \in \Phi \backslash \{ b_0 \} } {g_i R_i ^{ - \alpha } } } \right)} \right]\nonumber\\
&\mathop=\limits^{(a)}& \left( {{\mu  \over {\mu  + sr^{ - \alpha } }}} \right)^2 \mathbb{E}_\Phi  \left[ {\prod\limits_{i \in \Phi \backslash \{ b_0 ,b_1 ,b_2 \} } {\mathbb{E}_g \left[ {\frac{\mu}{\mu+sR_i^{-\alpha}} } \right]} } \right]
\nonumber\\
&\mathop=\limits^{(b)}&\left( {{\mu  \over {\mu  + sr^{ - \alpha } }}} \right)^2 \exp \left( { - 2\lambda \pi \int_r^\infty  {\left( {{{sv^{ - \alpha } } \over {\mu  + sv^{ - \alpha } }}} \right)} {\rm{ }}vdv} \right),\nonumber\\
\end{eqnarray}
where $(a)$ follows from the Rayleigh fading assumption (i.e., $g_i \sim \exp(\mu)$), and $(b)$ follows from the
probability generating functional (PGFL) \cite{Stoyan_96} of the
PPP. Note that Palm bias does not affect step (b) of (13) due to Lemma 1 in \cite{J_Meche_95}.
Putting $s=\mu\gamma r^{\alpha}$ and using a change of variables $u = v^2 \gamma ^{ - 2/\alpha } r^{ - 2}$ give
\setlength\arraycolsep{3pt}
\begin{eqnarray}
L_{I_r } \left( {\mu \gamma r^\alpha  } \right) = \left( {{1 \over {1 + \gamma }}} \right)^2 e^{ - \lambda \pi r^2 \gamma ^{2/\alpha } \int_{\gamma ^{ - 2/\alpha } }^\infty  {{1 \over {1 + u^{\alpha /2} }}du} }.
\end{eqnarray}
Then, the \textrm{P}roposition 1 is obtained by putting (14) into (12).

\subsection{Proof of Corollary 1}
For positive real-valued random variables $g_0,\ldots, g_n$, define
\begin{eqnarray}
G=max(g_0,\ldots, g_n)
\end{eqnarray}
Since the random variables $g_i, i=0,\ldots, n$ are i.i.d. exponential RVs with mean $1/\mu$,
the complementary cumulative density function (CCDF) of $G$ is
\begin{eqnarray}
\mathbb{P}\left( {G > g} \right) =1- {\left( {1 - e^{ - \mu g} } \right)}^{n+1}.
\end{eqnarray}

In the CS scheme, only one BS that has the best channel condition from the worst-case MU transmits to
it so that the coverage probability  is
\setlength\arraycolsep{1pt}
\begin{eqnarray}
p_c ^{CS} (\gamma ,\lambda ,4)=\int_{r > 0}^{} {\mathbb{P}\left[ {{{\max \left( {g_0 ,g_1 ,g_2 } \right)r^{ - 4} } \over {I_r^{CS} }} > \gamma |r} \right]} {\rm{ }}f_r (r)dr, \nonumber \\
\end{eqnarray}
where $I_r ^{CS}$ is the total interference from the all interfering BSs located farther than $r$.
For $n=2$ of (16), we have
\setlength\arraycolsep{-1pt}
\begin{eqnarray}
&&\mathbb{P}\left[ {\max \left( {g_0 ,g_1 ,g_2 } \right) > \gamma r^4 I_r ^{CS} |r} \right]\nonumber\\
&&=3\mathcal{L}_{I_r ^{CS} } \left( {\mu \gamma r^4 } \right) - 3\mathcal{L}_{I_r ^{CS} } \left( {2\mu \gamma r^4 } \right) + \mathcal{L}_{I_r ^{CS} } \left( {3\mu \gamma r^4 } \right).
\end{eqnarray}
Following the similar steps of (13) and (14), we achieve
\medmuskip=5mu
\begin{eqnarray}
&&3\mathcal{L}_{I_r ^{CS} } \left( {\mu \gamma r^4 } \right) - 3L_{I_r ^{CS} } \left( {2\mu \gamma r^4 } \right) + L_{I_r ^{CS} } \left( {3\mu \gamma r^4 } \right)\nonumber\\
&&=3e^{ - \lambda \pi r^2 \rho (\gamma ,4)}  - 3e^{ - \lambda \pi r^2 \rho (2\gamma ,4)}  + e^{ - \lambda \pi r^2 \rho (3\gamma ,4)}.
\end{eqnarray}
Putting (19) into (17) and using integration by parts give the desired result.

\subsection{Proof of Proposition 2}
Using the property $\mathbb{E}[X] = \int_{t > 0}^{}
{\mathbb{P}\left( {X > t} \right)dt}$ for a positive random variable
X, the spectral efficiency is
\setlength\arraycolsep{-1pt}
\begin{eqnarray}
&&\tau_c (\lambda ,4) \buildrel \Delta \over = \mathbb{E}\left[ {\ln
\left( {1
+SIR} \right)} \right]\nonumber \\
&&= \int_{r > 0}^{} {\int_{t > 0}^{} {\mathbb{P}\left[ {\ln \left(
{1 + {{g_0 r^{ - 4} } \over {I_r }}} \right) > t} \right]} {\rm{
}}dt{\rm{
}}f_r (r)dr}\nonumber \\
&&=\int_{r > 0}^{} {e^{ -  \lambda\pi r^2 } \int_{t > 0}^{} {\mathcal{L}_{I_r } \left( {\mu r^4 \left( {e^t  - 1} \right)} \right)} } {\rm{ }}dt2\pi ^2 \lambda ^2 r^3 dr.
\end{eqnarray}
From (13) and (14), we obtain
\setlength\arraycolsep{3pt}
\medmuskip=-1mu
\begin{eqnarray}
L_{I_r } \left( {\mu r^4 \left( {e^t  - 1} \right)} \right) = e^{ - 2t - \lambda \pi r^2 \rho \left( {\sqrt{e^t  - 1} , 4} \right)}.
\end{eqnarray}
Then, \textrm{P}roposition 2 can be found by plugging (21) into (20) and using integration by parts.

\subsection{Proof of Corollary 2}
The spectral efficiency in the CS scheme is
\setlength\arraycolsep{-1pt}
\begin{eqnarray}
&&\tau _c ^{CS} (\lambda ,4)\nonumber\\
&&\mathop=\int_{r > 0}^{} {\int_{t > 0}^{} {\mathbb{P}\left[ {\ln \left( {1 + {{\max (g_0 ,g_1 ,g_2 )r^{ - 4} } \over {I_r ^{CS} }}} \right) > t} \right]} {\rm{ }}dtf_r (r)dr}. \nonumber\\
\end{eqnarray}
For $n=2$ of (16), we have
\setlength\arraycolsep{-2pt}
\begin{eqnarray}
&&\mathbb{P}\left[ {\max (g_0 ,g_1 ,g_2 ) > \left( {e^t  - 1} \right)r^4 I_r ^{CS} } \right]=3\mathcal{L}_{I_r ^{CS} } \left( {\mu r^4 (e^t  - 1)} \right)\nonumber \\
&&-3\mathcal{L}_{I_r ^{CS} } \left( {2\mu r^4 (e^t  - 1)} \right) + \mathcal{L}_{I_r ^{CS} } \left( {3\mu r^4 (e^t  - 1)} \right).
\end{eqnarray}
Following the similar steps of (14) gives
\thinmuskip=-2mu
\medmuskip=-1mu
\thickmuskip=0mu
\setlength\arraycolsep{0pt}
\begin{eqnarray}
&&3\mathcal{L}_{I_r ^{CS} } \left( {\mu r^4 (e^t  - 1)} \right)-3\mathcal{L}_{I_r ^{CS} } \left( {2\mu r^4 (e^t  - 1)} \right) + \mathcal{L}_{I_r ^{CS} } \left( {3\mu r^4 (e^t  - 1)} \right)\nonumber\\
&&=3e^{ - \lambda \pi r^2 \rho \left( { {e^t  - 1},4} \right)}  - 3e^{ - \lambda \pi r^2 \rho \left( {2\left( {e^t  - 1} \right),4} \right)}  + e^{ - \lambda \pi r^2 \rho \left( {3\left( {e^t  - 1} \right),4} \right)}. \nonumber\\
\end{eqnarray}
Putting (24) into (22) and using integration by parts give the desired result.

\end{document}